\documentclass[12pt]{iopart}
\usepackage{epsfig}
\usepackage{dcolumn}
\usepackage{bm}
\usepackage{hyperref}
\usepackage{latexsym}

\begin{document}

\title{A possible classification of nonequilibrium steady states}
\author{R.~K.~P.~Zia and B.~Schmittmann}
\date{April 4, 2006}

\address{Center for Stochastic Processes in Science and Engineering and\\
Department of Physics, Virginia Tech, Blacksburg, VA 24061-0435 USA}

\begin{abstract}
We propose a general classification of nonequilibrium steady states in terms
of their stationary probability distribution and the associated probability
currents. The stationary probabilities can be represented
graph-theoretically as directed labelled trees; closing a single loop in such
a graph leads to a representation of probability currents. This
classification allows us to identify all choices of transition rates, based
on a master equation, which generate the same nonequilibrium steady state.
We explore the implications of this freedom, e.g., for entropy production.
\end{abstract}
\vspace{-0.8cm}
%PACS =============================
%      05.70.Ln  % Nonequilibrium and irreversible thermodynamics
%      02.50.Ga  % Markov processes%    
%      02.70.Rr  % General statistical methods
%      05.50.+q  % Lattice theory and statistics
\ead{\mailto{rkpzia@vt.edu}, \mailto{schmittm@vt.edu}}
%\vspace{2pc}
%\noindent{\it Keywords}: Nonequilibrium steady states, 
%     master equations, detailed balance, probability currents. \\
\submitto{\JPA} %PACS
%\vspace{-0.5cm}
%%%%%%%%%%%%%%%%%%%%%%%%%%%%%%%%%%%%%%%%%%%%%%%%%%%%%%%%%%%%%%%%%%%%%%%%%%%%%%%%%%%%%%%%%%%%%%%%%%%%%%%%%%%%%%%%%%%%%%%%%%%%

\emph{Introduction. }One of the greatest successes of statistical mechanics
was forging the fundamental link between microscopic interactions and
macroscopic behavior for interacting many-body systems in equilibrium.
Boltzmann and Gibbs established the general framework which allows us to
compute -- at least in principle -- any macroscopic observable of interest.
Labeling the microscopic states (``configurations'', $\mathcal{C}$) of the
system and establishing a form for the internal energy $\mathcal{H}(\mathcal{%
C})$ associated with each $\mathcal{C}$, macroscopic observables can be
expressed as statistical averages, $\left\langle A\right\rangle =$ $\sum_{%
\mathcal{C}}$ $A(\mathcal{C)}P^{eq}(\mathcal{C)}$, with the appropriate
equilibrium distribution $P^{eq}(\mathcal{C)}$.

Given that a real system continuously undergoes transitions from one
configuration to another, it is quite remarkable that $P^{eq}(\mathcal{C)}$
can be determined without explicit recourse to a time-dependent
distribution, $P(\mathcal{C};t\mathcal{)}$. At the root of this enormous
simplification lies the property of \emph{detailed balance}. Related to
microscopic reversibility, a system evolving according to a dynamics with
this property will eventually settle in a stationary state in which the 
\emph{net} probability current between \emph{any pair} of configurations
vanishes. As a result, $P^{eq}(\mathcal{C)\equiv }\lim_{t\rightarrow \infty
}P(\mathcal{C};t\mathcal{)}$ can be expressed in terms of \emph{ratios} of
the (dynamic) transition rates between configurations, and the long-time
limit remains invariant under any modification of the dynamics which
preserves these \emph{ratios}. Indeed, Monte Carlo simulation studies of
equilibrium systems rely heavily on this property.

In summary, systems in thermal equilibrium are fundamentally well
understood, including their dynamical representations. In stark contrast, a
comparable theoretical framework is still sorely lacking for systems far
from thermal equilibrium. Even the simplest generalizations of thermal
equilibrium, namely, non-equilibrium steady states (NESS), are currently
analyzed case-by-case. Thus, much effort is directed at simple models,
maintained far from equilibrium by imposing external driving forces, with
the goal of identifying some generic classes of NESS \cite{sz,Mukamel}.
Typically, these models are specified by a set of transition rates
(motivated by physical considerations), so that the master equation provides
a natural framework for analysis. A key feature of far-from-equilibrium
dynamics, these rates \emph{violate} detailed balance, resulting in \emph{%
non-vanishing }probability currents for the final, time-independent NESS. In
general, its stationary distribution is \emph{not }known a priori, and must
be found by solving the master equation. Thus, it is impossible to discern
if two different sets of rates will lead to the \emph{same} NESS without
solving both master equations. By contrast, for equilibrium systems with
microscopically reversible dynamics, simply comparing the ratios of the
rates will suffice!

In this letter, we address these fundamental issues. Starting from a \emph{%
general} master equation which admits a unique stationary distribution, $%
P^{*}(\mathcal{C})$, we first review a graphical construction for $P^{*}$ in
terms of (directed) labelled trees. Established some time ago \cite%
{Hill,Schn,HH}, this method seems not to be widely known. In this approach,
the non-trivial \emph{probability currents }associated with NESS are very
naturally associated with the violation of detailed balance. Second, we
propose a general classification of NESS in terms of \emph{both }$P^{*}(%
\mathcal{C})$ \emph{and }the (stationary)\ \emph{probability currents}, $%
K^{*}(\mathcal{C},\mathcal{C}^{\prime })$, between configurations $\mathcal{C%
}$ and $\mathcal{C}^{\prime }$. In other words, we postulate that a complete
description for a NESS is $\{P^{*},K^{*}\}$, being the appropriate
generalization of the Boltzmann distribution for equilibrium systems: $%
\{P^{eq},0\}$. In addition to the usual macroscopic averages $\left\langle
A\right\rangle $, $\{P^{*},K^{*}\}$ allows us to compute \emph{fluxes }of
all physical quantities (e.g., mass and energy currents), within and through
our system. In this framework, we can specify the class of transition rates
which lead to the same NESS, leading to a generalization of the ``detailed
balance condition'' routinely exploited in simulation studies of equilibrium
systems. In other words, all transformations of the rates which leave $%
\{P^{*},K^{*}\}$ invariant are known. Further, \emph{if} we are provided two
distinct set of rates corresponding to the same NESS, then $\{P^{*},K^{*}\}$
can be found trivially. We conclude with a discussion of entropy production
and some general comments.

\emph{The master equation and graphic representation of }$P^{\ast }$\emph{.}%
\ We begin with a generic master equation for an interacting many-body
system with a finite number ($N$) of configurations. Labelling the
configurations in some arbitrary fashion as $\mathcal{C}_{1}$, $\mathcal{C}%
_{2}$, .., $\mathcal{C}_{N}$, we write the transition rate, per unit time,
from $\mathcal{C}_{j}$ to $\mathcal{C}_{i}$ as $w_{i}^{j}$. All $w$'s are
real, non-negative, and assumed to be time-independent. In general, $%
w_{i}^{j}$ differs from its reverse, $w_{j}^{i}$. The master equation for $%
P_{i}(t)\equiv P(\mathcal{C}_{i},t)$, the probability to find the system in
configuration $\mathcal{C}_{i}$ at time $t$ reads: 
\begin{equation}
\partial _{t}P_{i}(t)=\sum_{j\neq i}\left[
w_{i}^{j}P_{j}(t)-w_{j}^{i}P_{i}(t)\right] \equiv \sum_{j}W_{i}^{j}P_{j}(t)
\label{me}
\end{equation}
i.e., the off- and on-diagonal elements of $W$ are just $w_{i}^{j}$ and $%
(-\sum_{\neq j}w_{i}^{j})$. Note that $\sum_{i}W_{i}^{j}=0$, for $\forall $ 
$j$, ensuring that $\sum_{i}P_{i}(t)=1$, for $\forall $ $t$. Such a $W$ is
known as a stochastic matrix. Since equation~(\ref{me}) is just a continuity
equation for probability, we can write $K_{i}^{j}(t)\equiv
w_{i}^{j}P_{j}(t)-w_{j}^{i}P_{i}(t)$ as the net \emph{probability} current
from $\mathcal{C}_{j}$ into $\mathcal{C}_{i}$.

In the following, we assume that every configuration can be reached from
every other configuration. Under these conditions, equation~(\ref{me}) is
ergodic and has a unique stationary solution, $P_{i}^{\ast }\equiv
\lim_{t\rightarrow \infty }P_{i}(t)$. The associated stationary currents are
denoted by $K^{\ast }{}_{i}^{j}$ and satisfy $\sum_{j\neq i}K^{\ast
}{}_{i}^{j}=0$, i.e., the total probability current into any given
configuration vanishes. If the rates satisfy detailed balance, as for
systems evolving towards thermal equilibrium where $%
w_{i}^{j}/w_{j}^{i}=P_{i}^{eq}/P_{j}^{eq}$, then all stationary currents%
\emph{\ vanish}. An equivalent statement of detailed balance which does not
reference $P_{i}^{eq}$ explicitly \cite{Mukamel} involves closed loops in
configuration space, i.e., $\mathcal{C}_{i}\rightarrow \mathcal{C}%
_{j}\rightarrow \mathcal{C}_{k}\rightarrow ...\rightarrow \mathcal{C}%
_{n}\rightarrow \mathcal{C}_{i}$. For each loop, we define the product of
the associated rates in the ``forward'' and the ``reverse'' directions: $\Pi
_{f}\equiv w_{j}^{i}w_{k}^{j}...w_{i}^{n}$ and $\Pi _{r}\equiv
w_{i}^{j}w_{j}^{k}...w_{n}^{i}$. The detailed balance condition corresponds
to $\Pi _{f}=\Pi _{r}$ for all loops. Related to integrability, this
property allows $P_{i}^{\ast }$ to be computed from the $w$'s easily.

%===============================
\begin{figure}[tbp]
\begin{center}
\epsfig{file=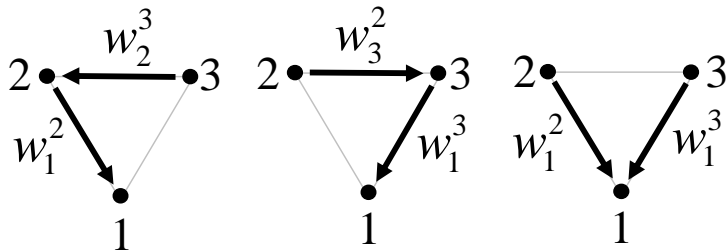,width=4.0in}
\end{center}
\par
\vspace{-0.4cm}
\caption{Representation of $P^*_{1}$, for a simple model with $N=3$, in
terms of directed labelled trees.}
\label{fig:fig-1}
\end{figure}
%===========================

In the absence of detailed balance, $P_{i}^{\ast }$ has to be found, in
principle, from $\sum_{j}W_{i}^{j}P_{j}^{\ast }=0$ laboriously. Fortunately,
there is a systematic way to construct $P^{\ast }$ \cite{Hill,Schn}, using
graph theoretical methods similar to those originally designed for electric
networks \cite{Kirchhoff}. First, associate each $\mathcal{C}_{i}$ with a
vertex, labelled $i$. Next, consider all distinct labelled trees (i.e., a
graph consisting of all vertices with a single undirected edge between each
pair, forming no loops). Denoting these trees as $t_{\alpha }$, $\alpha
=1,2,..,M$, Cayley's theorem \cite{Cayley} states that $M=N^{N-2}$. To
compute $P_{i}^{\ast }$, \emph{direct every edge} towards the vertex $i$ and
denote this subset of (directed)\ trees by $t_{\alpha (i)}$. In other words, 
$\alpha (i)$ runs over the set of directed trees with $i$ as the ``root''.
Next, a factor of $w_{k}^{n}$ is associated with an edge directed from $n$
to $k$. Finally, to each tree $t_{\alpha (i)}$, we assign a numerical value, 
$U(t_{\alpha (i)})$, which is the product of the $(N-1)$ factors of $w$'s in
the tree. Clearly, $U(t_{\alpha (i)})=0$ if one of the associated rates
vanishes. Then, the stationary distribution is given by 
\begin{equation}
P_{i}^{\ast }=\mathcal{Z}^{-1}\sum_{\alpha (i)}U(t_{\alpha (i)})
\label{P-star}
\end{equation}
where $\mathcal{Z}$ is just the normalization factor and may play the role
of a (super-) partition function. We illustrate the procedure in figure 1
for $N=3$. A more detailed discussion, including further examples, can be
found in \cite{trees-long}.

\emph{Probability currents and loops in configuration space.} From the
defining equation for $K_{i}^{j}$ above, we arrive at the \emph{net }%
(stationary)\emph{\ }probability current, from $\mathcal{C}_{j}$ into $%
\mathcal{C}_{i}$:

\begin{equation}
K^{\ast }{}_{i}^{j}\equiv w_{i}^{j}P_{j}^{\ast }-w_{j}^{i}P_{i}^{\ast }=%
\mathcal{Z}^{-1}\sum_{\alpha} \left[ w_{i}^{j}U(t_{\alpha (j)})-w_{j}^{i}
U(t_{\alpha(i)})\right]  \label{J-star}
\end{equation}

Focusing on the expression within $[...]$, we note that, for a specific $%
\alpha $, the trees $t_{\alpha (i)}$ and $t_{\alpha (j)}$ differ only in the
directed edges that connect vertices $i$ and $j$ (figure 2). Now,
multiplication of $U(t_{\alpha (j)})$ by $w_i^j$ can be regarded as adding a
directed edge from $j$ to $i$, converting $t_{\alpha (j)}$ into a graph with
a \emph{single} loop. Associated with this loop is the product $\Pi
_i^j(t_{\alpha (j)})\equiv w_i^j\left( w_{k_1}^iw_{k_2}^{k_1}...w_j^{k_\ell
}\right) $, where $k_1,...,k_\ell $ label the vertices between $i$ to $j$.
Similar considerations for $w_j^iU(t_{\alpha (i)})$ lead to a graph with the 
\emph{same} loop, but traversed in the \emph{opposite} sense and so,
associated with $\Pi _j^i(t_{\alpha (i)})=w_j^i\left( w_{k_\ell
}^j...w_i^{k_1}\right) $. Meanwhile, the rest of both trees (the side
branches of the loops) are identical, so that $R(t_{\alpha (i)})=R(t_{\alpha
(j)})$, where $R$ denotes the products of the $w$'s in the side branches.
Summarizing, we write $w_j^iU(t_{\alpha (i)})=\Pi _j^i(t_{\alpha
(i)})R(t_{\alpha (i)})$, etc., so that 
\begin{equation}
K^{*}{}_i^j=\mathcal{Z}^{-1}\sum_{\alpha} \left[ \Pi _i^j(t_{\alpha (j)})-\Pi
_j^i(t_{\alpha (i)})\right] R(t_{\alpha (i)})  \label{J-star-loops}
\end{equation}
This expression explicitly demonstrates the emergence of nonzero steady
state probability currents from rates which violate detailed balance, as
manifested in irreversible loops.

%===============================
\begin{figure}[tbp]
\begin{center}
\epsfig{file=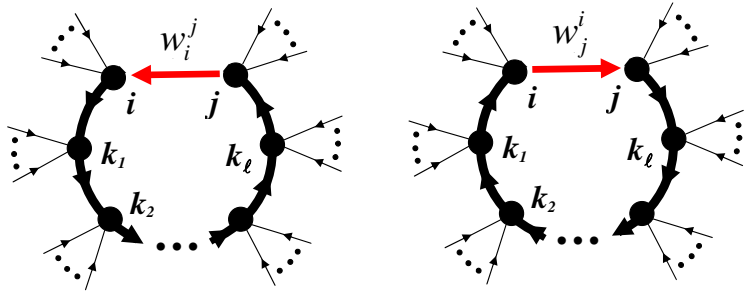,width=4.0in}
\end{center}
\par
\vspace{-0.4cm}
\caption{One term in $\left[ ...\right] $ of equation (\ref{J-star}),
illustrating the emergence of nontrivial loops. Straight arrows (red 
online) represent the $w$ factors; the rest (black online), the $U$'s.}
\label{fig:fig-2}
\end{figure}
%===========================

\emph{A postulate. }It is easy to construct cases where two sets of rates,
one obeying detailed balance, and the other violating it, lead to the same
stationary distribution. For example, for particles hopping on a ring with
symmetric or biased rates, $P^{eq}=P^{\ast }\propto 1$. Hence, one might be
tempted to consider any $P^{\ast }$ as an ``effective'' equilibrium system.
Indeed, nothing prevents us from labeling $-\ln P^{\ast }(\mathcal{C})$ as
an ``effective Hamiltonian''. Yet, in the simple example above, the system
with biased rates carries a physical current while the system with symmetric
rates does not. Thus, it is essential to identify a key signature that
distinguishes a NESS from an equilibrium state. We believe that the
nonvanishing stationary currents, $K^{\ast }$, fill this role. Therefore, we
propose that $\{P^{\ast },K^{\ast }\}$ form a \emph{complete and unique}
description for \emph{any} stationary state. The major difference between a
NESS and the standard equilibrium case is $K^{\ast }$ being nonzero. In this
sense, the class of NESS is significantly broader than equilibrium states,
and their analogs in electrodynamics would be, respectively, magnetostatics
and electrostatics. Given $\{P^{\ast },K^{\ast }\}$, all macroscopic
stationary properties of the system can be computed. Beyond the usual
averages with $P^{\ast }$ as the weights, the $K^{\ast }$ generate average 
\emph{fluxes }(or currents) associated with physical observables, such as
energy or particle number density. More details and examples will be
discussed elsewhere \cite{trees-long}.

\emph{Dynamic equivalence classes.} A common notion of NESS is that, unlike
their counterparts in thermal equilibrium, seemingly slight modifications of
the rates lead to very different steady states. Armed with our
characterization of a NESS in terms of $\{P^{*},K^{*}\}$, we can pose a
natural question: What determines the class of dynamics that leads to the
same NESS? In other words, given a set of $w$'s and its associated NESS,
what are the transformations (on the rates) which leave $\{P^{*},K^{*}\}$
invariant? For the equilibrium case, $\{P^{eq},0\}$, the answer is well
known:\ Any set of $w$'s will lead to a desired $P^{eq}$, provided they
satisfy the ``detailed balance condition'': $w_i^j/w_j^i=P_i^{eq}/P_j^{eq}$.
This can be regarded as a constraint on the $w$'s, \emph{given }a certain $%
P^{eq}$ is to be achieved. In our framework, this constraint can now be
easily generalized: To arrive at a \emph{given }$\{P^{*},K^{*}\}$ final
state, the $w$'s must satisfy \ 
\begin{equation}
w_i^jP_j^{*}-w_j^iP_i^{*}=K^{*}{}_i^j\,\,.  \label{w-rel}
\end{equation}
for all pairs $i\neq j$. In the remainder of this letter, we will explore
other representations of this constraint and some of its implications.

Let us decompose $WP^{\ast }$ into its symmetric and antisymmetric parts: 
\begin{equation}
W_{i}^{j}P_{j}^{\ast }=S_{i}^{j}+A_{i}^{j}  \label{def-M}
\end{equation}%
where $S\equiv (WP^{\ast }+WP^{\ast })/2$, and $A\equiv (WP^{\ast }-WP^{\ast
})/2$. Then equation (\ref{w-rel}) is just the constraint that $A$ is fixed
to be $K^{\ast }/2$. In contrast, there is no such constraint on $S$, except
for two restrictions. The physical rates must be non-negative ($w\geq 0$),
leading to $S_{i}^{j}\geq |A_{i}^{j}|$, $\forall i\neq j$. Next, probability
conservation imposes $\sum_{i}S_{i}^{j}=0$, $\forall $ $j$. Thus, we arrive
at the conditions 
\begin{equation}
S_{i}^{j}\geq \frac{1}{2}\left| K^{\ast }{}_{i}^{j}\right| \quad \forall
i\neq j,\qquad S_{j}^{j}=-\sum_{i\neq j}S_{i}^{j}\,.  \label{M-const}
\end{equation}%
Within these constraints, we can choose \emph{arbitrary} $S$'s and construct
the associated transition rates via 
\begin{equation}
W_{i}^{j}=\left[ S_{i}^{j}+\frac{1}{2}K^{\ast }{}_{i}^{j}\right]
(P_{j}^{\ast })^{-1}\quad ,  \label{new-W}
\end{equation}%
resting assured that the final NESS \emph{will remain the same}. Thus, we
may associate such $S$'s with an ``equivalence class'' of dynamical rates
leading to the same given NESS.

It is very instructive to consider the difference, $\Delta _{i}^{j}$,
between two sets of rates that belong to the same class. Since the
differences in the $S$'s are symmetric, $\Delta $ must satisfy 
\begin{equation}
\Delta _{i}^{j}P_{j}^{\ast }=\Delta _{j}^{i}P_{i}^{\ast }\,\,.
\label{NESS-DB}
\end{equation}
Reminiscent of the ordinary detailed balance condition, it is possible to
turn this into a mnemonic: ``The \emph{differences} (as opposed to the rates
themselves) must satisfy detailed balance with respect to $P^{\ast }$.''
Finally, we note an implication of this curious equation: If \emph{two} sets
of rates are somehow known to generate the same NESS, their (non-vanishing)
differences will provide a simple route to finding $P^{\ast }$. In contrast,
for equilibrium states, the stationary distribution can be easily generated
from just \emph{one }set of rates, again due to the \emph{known }absence of $%
K^{\ast }$'s in this case.

\emph{Entropy production. }One of the key signatures of nonequilibrium
steady states, recognized over three decades ago \cite{GP,Schn,Sei}, is
entropy production. For a general time-dependent solution of the master
equation, two independent quantities were introduced: the entropy production
of the ``system'' and of the ``medium'', 
\begin{equation}
\dot{\mathbf{S}}_{sys}\equiv \sum_{i,j}W_{i}^{j}P_{j}(t)\ln \frac{P_{j}(t)}{%
P_{i}(t)}\,\,,\,\,\,\dot{\mathbf{S}}_{med}\equiv
\sum_{i,j}W_{i}^{j}P_{j}(t)\ln \frac{W_{i}^{j}}{W_{j}^{i}}\,.
\label{S-sys+med}
\end{equation}%
The former is readily recognized as the time derivative of $\mathbf{S}%
_{sys}\equiv -\sum_{i}P_{i}(t)\ln P_{i}(t)$, which motivates the term
``entropy production of the system''. The latter is attributed to the
coupling of the system to the \emph{external environment} in a manner that
prevents it from reaching equilibrium \cite{Schn}. Neither $\dot{\mathbf{S}}%
_{sys}$ nor $\dot{\mathbf{S}}_{med}$ is necessarily positive. However,
their sum, naturally termed the ``total entropy production''\ 
\begin{equation}
\dot{\mathbf{S}}_{tot}\equiv \sum_{i,j}W_{i}^{j}P_{j}(t)\ln \frac{%
W_{i}^{j}P_{j}(t)}{W_{j}^{i}P_{i}(t)}\,\,,  \label{S-tot}
\end{equation}%
is indeed \emph{non-negative} \cite{Schn}.

Recasting these expressions in terms of the probability currents \cite{Schn}%
, and taking $t\rightarrow \infty $ to focus on stationary states, we arrive
at $\dot{\mathbf{S}}_{sys}^{\ast }=\frac{1}{2}\sum_{i,j}K^{\ast
}{}_{i}^{j}\ln \left( P_{j}^{\ast }/P_{i}^{\ast }\right) $ and $\dot{%
\mathbf{S}}_{med}^{\ast }=\frac{1}{2}\sum_{i,j}K^{\ast }{}_{i}^{j}\ln
(W_{i}^{j}/W_{j}^{i})$. For equilibrium states, both trivially vanish since
all currents are zero. By contrast, $K^{\ast }\neq 0$ for a NESS, though $%
\dot{\mathbf{S}}_{sys}^{\ast }$ remains zero (so that $\dot{\mathbf{S}}%
_{med}^{\ast }=\dot{\mathbf{S}}_{tot}^{\ast }$). The interpretion of these
results is clear: In the steady state, the entropy ``associated with our
system'' no longer changes. However, it is reasonable to expect that, being
coupled in an irreversible way to the environment, such a NESS continues to
``induce'' the entropy of its surrounding medium to increase (indeed, $\dot{%
\mathbf{S}}_{med}^{\ast }=\dot{\mathbf{S}}_{tot}^{\ast }>0$). In this
sense, $\dot{\mathbf{S}}_{med}^{\ast }$ carries detailed information of
transition rates and so, the precise nature of the coupling between our
system and its environment. As a result, even if we insist on having the
same NESS (i.e., a given $\{P^{\ast },K^{\ast }\}$), $\dot{\mathbf{S}}%
_{med}^{\ast }$ will not be unique. Let us explore the implications of these
``degrees of freedom''.

Since $\dot{\mathbf{S}}_{med}^{\ast }=\dot{\mathbf{S}}_{tot}^{\ast }$, we
focus on the latter and exploit equation (\ref{def-M}):\ 
\begin{equation}
\dot{\mathbf{S}}_{tot}^{\ast }=\frac{1}{2}\sum_{i,j}K^{\ast
}{}_{i}^{j}{}\ln \frac{S_{i}^{j}+A_{i}^{j}}{S_{i}^{j}-A_{i}^{j}}\,\,.
\label{S-tot-J}
\end{equation}%
Since $A=K^{\ast }/2$, the freedom we have is \emph{any} $S$ satisfying
equation (\ref{M-const}). An immediate consequence is that rates can be
chosen to \emph{minimize} the entropy production (associated with a given
NESS), by having $S\gg A$. To lowest order in $K^{\ast }{}_{i}^{j}/S_{i}^{j}$%
, we have $\dot{\mathbf{S}}_{tot}^{\ast }=\sum \left( K^{\ast }\right)
^{2}/\left( 2S\right) $. Since $\dot{\mathbf{S}}_{med}^{\ast }\equiv 0$ for
equilibrium cases, we can choose rates which are arbitrarily
``equilibrium-like'' in this respect. We should emphasize that, though $\dot{%
\mathbf{S}}_{tot}^{\ast }$ can be made arbitrarily small, it remains
strictly positive and retains the NESS\ signature. At the opposite extreme,
we can consider rates with ``infinite $\dot{\mathbf{S}}_{med}^{\ast }$'' by
lowering some $S_{i}^{j}$ to $\left| A_{i}^{j}\right| $. Whether such a
concept is useful deserves further exploration. It is natural to label such
rates as ``maximally asymmetric'', since one of the two directed edges
between some pairs of configurations is \emph{missing}. Such models abound
in the literature, e.g., totally asymmetric exclusion processes (TASEP) \cite%
{TASEP}. One clear advantage of having maximally asymmetric rates for all
edges is that the number of trees used for constructing $P^{\ast }$ is kept
at the absolute minimum. Of course, the expression for $K^{\ast }$ also
simplifies. Finally, the implications for non-equilibrium work theorems \cite%
{Jar,Sei} are not trivial and will be published elswhere \cite{trees-long}.

\emph{Conclusions.} To summarize, we have addressed a fundamental question
associated with non-equilibrium steady states: Within the framework of the
master equation, what class, if any, of transition rates $W$ lead to the
same stationary state? For equilibrium systems, the answer is provided by
the detailed balance condition. To generalize this answer to NESS, we first
postulate that a NESS is completely and uniquely specified by its stationary
distribution $P^{\ast }$ \emph{in conjunction with} the steady currents $%
K^{\ast }$. Then the \emph{generalized} detailed balance condition is simply
equation (\ref{w-rel}). Exploiting a graphic method to compute $P^{\ast }$
in terms of directed labelled trees, we display the connection between $%
K^{\ast }$ and ``irreversible'' loops - key characteristics of rates that
violate detailed balance. Extensions, examples, and further implications of
these explorations may be found in \cite{trees-long}.

\textbf{Acknowledgements.} We thank U.~Seifert for fruitful discussions. BS
acknowledges the hospitality of the Isaac Newton Institute in Cambridge, UK,
where some of this work was performed. Financial support from the NSF
through DMR-0414122 is gratefully acknowledged.

\end{document}